\documentclass[amsmath,amssymb,aip,apl,reprint]{revtex4-1}

\usepackage{color}
\usepackage{graphicx}
\usepackage{subfigure}
\usepackage{verbatim}
\usepackage{fourier}
\usepackage{amsfonts}
\usepackage{multirow}
\usepackage{diagbox}
\usepackage{booktabs}
\usepackage[colorlinks,citecolor=blue,linkcolor=red,urlcolor=blue]{hyperref}

\begin{document}

\preprint{HIT-L2C-UM-CNRS}

\title{Near-field radiative heat transfer between twisted nanoparticle gratings}

\author{Minggang Luo}
\affiliation{School of Energy Science and Engineering, Harbin Institute of Technology, 92 West Street, Harbin 150001, China}
\affiliation{Laboratoire Charles Coulomb (L2C) UMR 5221 CNRS-Universit\'e de Montpellier, F- 34095 Montpellier, France}

\author{Junming Zhao}
\email[]{jmzhao@hit.edu.cn}
\affiliation{School of Energy Science and Engineering, Harbin Institute of Technology, 92 West Street, Harbin 150001, China}
\affiliation{Key Laboratory of Aerospace Thermophysics, Ministry of Industry and Information Technology, Harbin 150001, China}

\author{Mauro Antezza}
\email[]{mauro.antezza@umontpellier.fr}
\affiliation{Laboratoire Charles Coulomb (L2C) UMR 5221 CNRS-Universit\'e de Montpellier, F- 34095 Montpellier, France}
\affiliation{Institut Universitair\'e de France, 1 rue Descartes, F-75231 Paris Cedex 05, France}

\date{\today}

\begin{abstract}
We study the near-field radiative heat transfer between two twisted finite-size polar dielectric nanoparticle gratings. Differently from previous studies of the same configuration, we do not rely on any approximated effective medium theory to describe the gratings. By the full many-body radiative heat transfer theory we are able to investigate how the size, distance and relative orientation between the gratings influence the radiative heat flux. By changing the twisting angle $\theta$, we show a significant oscillation of the thermal conductance $G(\theta)$, due to the size effect for gratings of both square and circular shapes.  The distance- and twisting-dependent coupling between the gratings accounts for a strong and characteristic modulation of radiative thermal conductance with implications for the energy management, sensing, and NEMS/MEMS devices.
\end{abstract}

\maketitle

Near-field radiative heat transfer (NFRHT) has recently attracted much attention for both fundamental and applicative reasons. The radiative heat exchange between two objects with a separation distance comparable to or less than the thermal wavelength $\lambda_T=\hbar c/k_BT$ can exceed by several orders of magnitude the Planckian blackbody limit, and has been theoretically investigated for physical systems of different geometries (e.g., two planar surfaces \cite{Rytov1989,Polder1971,Loomis1994,Carminati1999,Shchegrov2000,Volokitin2001}, two isolated nanoparticles \cite{Chapuis2008,Manjavacas2012,Nikbakht2018}, two spheres \cite{Narayanaswamy2008}, one dipole and surface \cite{Chapuis2008plate}, two nanoparticles above a substrate \cite{Messina2018,DongPrb2018,Zhang2019R,He2019APL} and two nanoparticles separated by a multilayer plate \cite{Zhang2019T}) and has also been proved experimentally recently (e.g., two plates \cite{Ottens2011,Lim2015,Watjen2016,Ghashami2018,DeSutter2019}, one plate and one sphere or tip \cite{Shen2009,Rousseau2009,Song2015}).

  Active tuning of NFRHT is of great interest and importance for micro-nanoscale heat management. To this purpose, several proposals investigated the NFRHT between twisted gratings \cite{Biehs2011gratings}, possibly realised with graphene-coated strips \cite{He2020OL}, or using natural anisotropic materials (two multiple black phosphorus layers) with a twisting relative angle between the upper and lower multiple layers \cite{Zhang2018ACS}.  Some other important progresses have been also reported on tuning and manipulating the NFRHT in the micro-nano cale, e.g., pattern-free thermal modulator \cite{Liu2017vdw}, strain-induced modulation \cite{Ghanekar2018apl}, thermal routing \cite{Song2020routing}, thermal Hall effect \cite{Ben2016Hall}, heat flux splitter \cite{Ben2015splitter}, to name a few.

NFRHT modulations via twisted gratings typically use the effective medium approximation (EMA) \cite{Biehs2011OE,Tao1990,Haggans1993EMA} to describe the gratings, and consider the approximation of infinite-size systems \cite{Biehs2011gratings}. When the distance between two parallel gratings comparable to or less than the grating periods, the  EMA theory cannot be used. In addition, the finite size effect will also bring new challenges when using the EMA to calculate NFRHT between two finite gratings. For the finite size physical system, as compared to the corresponding infinite system, some new physical insight may be introduced. This has been recently shown when considering the Casimir torque between two finite gratings at different twisting angles, where the torque per area can reach extremely large values, increasing without bounds with the size of the system \cite{Mauro2020prl}. An investigation of finite-size and a beyond-EMA approximations effects between twisted gratings is still missing for the NFRHT.
 
\begin{figure} [htbp]
\centerline {\includegraphics[width=0.45\textwidth]{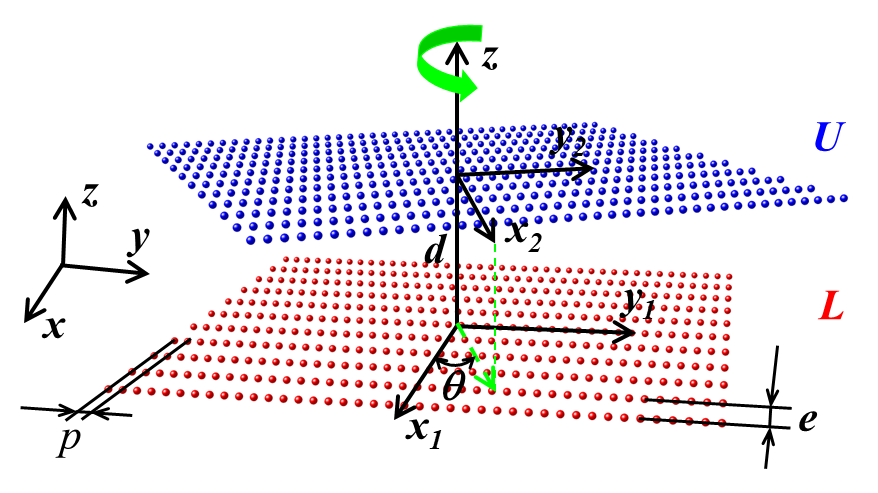}}
\caption{Two nanoparticle gratings (L and U), realized with parallel nanoparticle chains, are separated by a distance $d$ and twisted by an angle $\theta$. The grating lattice period is $e$, while $p$ is the distance between neighboring nanoparticles in each nanoparticle chain.}
\label{Structure_diagram}
\end{figure}

In this letter, by investigating the the NFRHT between two finite-size polar dielectric nanoparticle gratings with a relative twisting angle $\theta$ (see Fig.~\ref{Structure_diagram}) and by using a full many-body radiative heat transfer theory \cite{Ben2011}, we show that finite-size effects and an exact calculation beyond the EMA theory lead to new qualitative and quantitative affects and features.

As shown in Fig.~\ref{Structure_diagram}, each ensemble is composed of many nanoparticle chains. When calculating the radiative thermal conductance $G$, we consider the two ensembles are near to the thermal equilibrium around the temperature $T$. Inside each nanoparticle chain, $p$ is the distance between neighboring nanoparticles, while $e$ is the separation distance between two neighboring chains. Nanoparticle radius $a$ is fixed as the smallest length-scale in the problem, allowing for the dipole approximation \cite{Ben2019,DongPrb2017}. The separation distance between the gratings L and U center to center is $d$. The minimum nanoparticle separation in each chain is set to $p=3a$. When considering heat exchange between such nanoparticle ensembles where $e> 3p$, each ensemble behaves like a grating. In addition, the effects of the coupling between the two gratings, breaking of symmetry and collective many-body interaction on NFRHT between the two nanoparticle gratings is fully taken into account by the many-body radiative heat transfer theory \cite{Ben2013,Tervo2017,Tervo2019,DongPrb2017,Luo2019,Luo2019JQ}.

We focus on the radiative thermal conductance $G(\theta)$ between two polar dielectric SiC nanoparticle gratings with a relative angle $\theta$, which is sum of that between all possible nanoparticle pairs (one from the grating L and the other one from grating U) and is defined as follows \cite{Luo2020}:
\begin{equation}
G(\theta)=\sum_{\rm i\in U}^{}\sum_{\rm j\in L}^{} G_{\rm ij},
\label{Gt}
\end{equation}
where $G_{\rm ij}$ is the thermal conductance between two arbitrary nanoparticles i and j and yields
\begin{equation}
G_{\rm ij}=3\int_{0}^{+\infty} \frac{\mathrm{d}\omega}{2\pi}\frac{\partial\Theta(\omega,T)}{\partial T}\mathcal{T}_{\rm i,j}(\omega) ,
\label{Gij}
\end{equation}
where $\omega$ is angular frequency, $\Theta(\omega,T)$ is the mean energy of the Planck oscillator, transmission coefficient $\mathcal{T}_{\rm i,j}(\omega)$ between the jth and ith dielectric particles is given as follows \cite{Ben2011,DongPrb2017}.
\begin{equation}
\mathcal{T}_{\rm i,j}(\omega)=\frac{4}{3}k^4{\rm Im}\left(\chi_E^{\rm i}\right){\rm Im} \left(\chi_E^{\rm j}\right)\textrm{Tr}\left(G_{\rm ij}^{EE}G_{\rm ij}^{EE\dagger}\right),
\label{transmission}
\end{equation}
where the parameter $\chi_E^{\rm i~or~j}=\alpha_E^{\rm i~or~j}-\frac{ik^3}{6\pi}
\left|\alpha_E^{\rm i~or~j}\right|^2$ is introduced \cite{Manjavacas2012}, $\alpha_E^{\rm i~or~j}$ is the corresponding electric dipole polarizability described as $4\pi a^{3}(\varepsilon-1)/(\varepsilon+2)$ in the Clausius-Mossotti form\cite{Ben2011}, the permittivity for the polar dielectric SiC is described by the Drude-Lorentz model \cite{Palik} $\epsilon(\omega) =\epsilon_{\infty}^{}(\omega^2-\omega_l^2+i\gamma\omega)/(\omega^2-\omega_t^2+i\gamma\omega)$ with parameters $\epsilon_{\infty}^{}$ = 6.7, $\omega_l^{}$ = 1.827 $\times$ 10$^{14}$ rad$\cdot$s$^{-1}$, $\omega_t^{}$ = 1.495 $\times$ 10$^{14}$ rad$\cdot$s$^{-1}$, and $\gamma$ = 0.9 $\times$ 10$^{12}$ rad$\cdot$s$^{-1}$ , $k$ is vacuum wavevector, $G_{\rm ij}^{EE}$ is the electric-electric Green\textquotesingle s function in the particulate system considering many-body interaction, which is the element of the following left matrix.
\begin{equation}
\begin{pmatrix}
0 & G_{12}^{EE} & \cdots & G_{1N}^{EE}\\
G_{21}^{EE} & 0 & \ddots & \vdots\\
\vdots & \vdots & \ddots & G_{(N-1)N}^{EE}\\
G_{N1}^{EE} & G_{N2}^{EE} & \cdots & 0
\end{pmatrix}\mathbb{A}=\begin{pmatrix}
0 & G_{0,12}^{EE} & \cdots & G_{0,1N}^{EE}\\
G_{0,21}^{EE} & 0 & \ddots & \vdots\\
\vdots & \vdots & \ddots & G_{0,(N-1)N}^{EE}\\
G_{0,N1}^{EE} & G_{0,N2}^{EE} & \cdots & 0
\end{pmatrix},
\label{double_sources}
\end{equation}
where $G_{\rm 0,ij}^{EE}=\frac{e^{ikr}}{4\pi r}\left[\left(1+\frac{ikr-1}{k^{2}r^{2}}\right)\mathbb{I}_{3}+\frac{3-3ikr-k^{2}r^{2}}{k^{2}r^{2}}\hat{\textbf{r}}\otimes\hat{\textbf{r}}\right]$ is the free space Green\textquotesingle s function connecting two nanoparticles at $\textbf{r}_{\rm i}^{}$ and $\textbf{r}_{\rm j}^{}$, $r$ is the magnitude of the separation vector $\textbf{r}=\textbf{r}_{\rm i}^{}-\textbf{r}_{\rm j}^{}$, $\hat{\textbf{r}}$ is the unit vector $\textbf{r}/r$, $\mathbb{I}_{3}$ is a $3\times3$ identity matrix  and the matrix $\mathbb{A}$ including many-body interaction is defined as
\begin{equation}
\mathbb{A}=\mathbb{I}_{3N}^{}-k^2\begin{pmatrix}
0 & \alpha_{E}^{1}G_{0,12}^{EE} & \cdots & \alpha_{E}^{1}G_{0,1N}^{EE}\\
\alpha_{E}^{2}G_{0,21}^{EE} & 0 & \ddots & \vdots\\
\vdots & \vdots & \ddots & \alpha_{E}^{N-1}G_{0,(N-1)N}^{EE}\\
\alpha_{E}^{N}G_{0,N1}^{EE} & \cdots & \alpha_{E}^{N}G_{0,N(N-1)}^{EE} & 0
\end{pmatrix},
\label{matrix_interaction}
\end{equation} 
where $\mathbb{I}_{3N}$ is a $3N\times 3N$ identity matrix.

\begin{figure} [htbp]
\centerline {\includegraphics[width=0.5\textwidth]{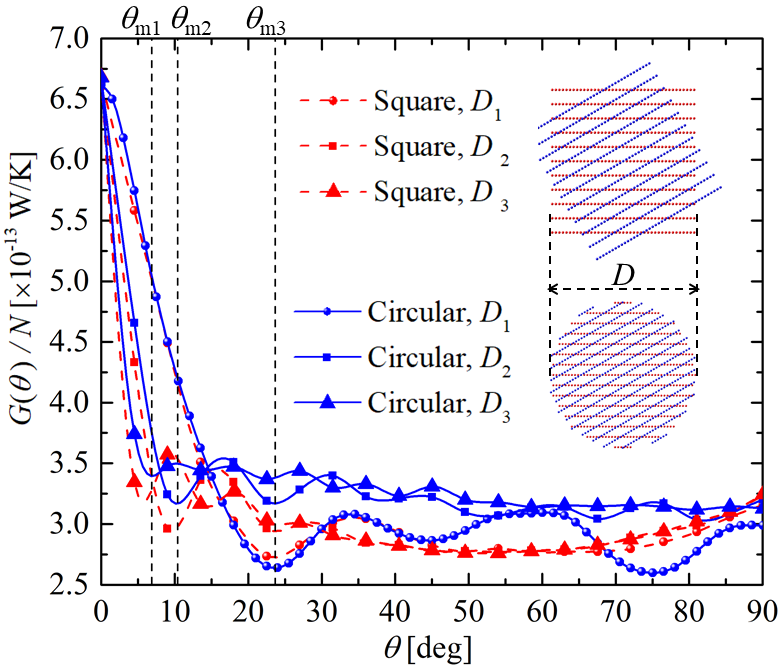}}
\caption{The scaled thermal conductance $G(\theta)/N$ as a function of the twisting angle $\theta$. A scheme for the square and circular grating configurations is shown in the inset. Three different lateral lengths are considered, $D_1^{}=0.8~\mu$m, $D_2^{}=1.4~\mu$m and $D_3^{}=2.8~\mu$m, respectively. $d=80$ nm, $a=20$ nm, $T=300$~K, $p=60$ nm and $e=10a=200$ nm. For the circular grating configuration, $\theta_{\rm m1}$, $\theta_{\rm m2}$ and $\theta_{\rm m3}$ correspond to the angle where $G(\theta)$ decreases to its first valley value, respectively  for systems having lateral lengths $D_1, D_2$ and $D_3$.}
\label{size_effect}
\end{figure}

Now, we discuss the numerical results using above expressions for NFRHT between two polar dielectric SiC nanoparticle gratings. In Fig.~\ref{size_effect}, we show the scaled thermal conductance $G(\theta)/N$ as a function of the twisting angle $\theta$. Both square gratings and circular gratings are considered, of which the schematics are shown in the inset of Fig.~\ref{size_effect}. Three lateral sizes are considered, $D_1^{}=0.8~\mu$m, $D_2^{}=1.4~\mu$m and $D_3^{}=2.8~\mu$m, respectively. $a=20$ nm, $T=300$ K, $d=80$ nm, $p=60$ nm and $e=$~200 nm.

For both of the two square and circular gratings, with increasing twisting angles, $G(\theta)$ decreases to its first valley and then oscillates slightly. The twisting angle corresponding to the first valley of the $G(\theta)$ is $\theta_{\rm m}$. For the circular gratings of three lateral lengths ($D_1^{}$, $D_2^{}$ and $D_3^{}$), $\theta_{\rm m}$ are $24^\circ$, $10^\circ$ and $6.8^\circ$, respectively. The $\theta_{\rm m}$ increases with decreasing the lateral length of the gratings. The amplitude of the $G(\theta)$ oscillation corresponding to $D_1^{}$ is much bigger than that corresponding to $D_2^{} ~{\rm and}~ D_3^{}$. As the lateral length of the circular grating increases, the amplitude of the oscillation of $G(\theta)$ decreases gradually. The thermal conductance $G(\theta)$ as a function of $\theta$ tends to converge to a given curve when the size of gratings increases. The size effect accounts for the oscillating behaviour of $G(\theta)$ and was absent in previous studies using infinite-systems and EMA approximations. It is worth stressing that an oscillation of the thermal conductance due to a lateral shift of two aligned gratings has been recently reported \cite{Luo2020}.

Even for the same value of the size $D$, the curves of $G(\theta)$ are quite different if ce consider square or circular gratings.
In particular,  for large $\theta$ (50$^\circ \sim 90^\circ$), $G(\theta)$ increases monotonically with $\theta$ for the square gratings, while the oscillations are still present for circular gratings.  Hence we deduce that the overlap between the L and U gratings, which is different in case of square and circular configurations, significantly influences the thermal conductance between two 2D ensembles, as it happens for the Casimir torque \cite{Mauro2020prl}. At large angles, the overlap area of the two square gratings increases monotonically with increasing the twisting angle $\theta$, which may account for the monotonically increasing dependence of $G(\theta)$ with $\theta$ shown by square gratings. For circular gratings, the overlap area will not vary with increasing $\theta$ and the angle dependent coupling accounts for the oscillation of $G(\theta)$ with $\theta$.

\begin{figure} [htbp]
\centerline {\includegraphics[width=0.48\textwidth]{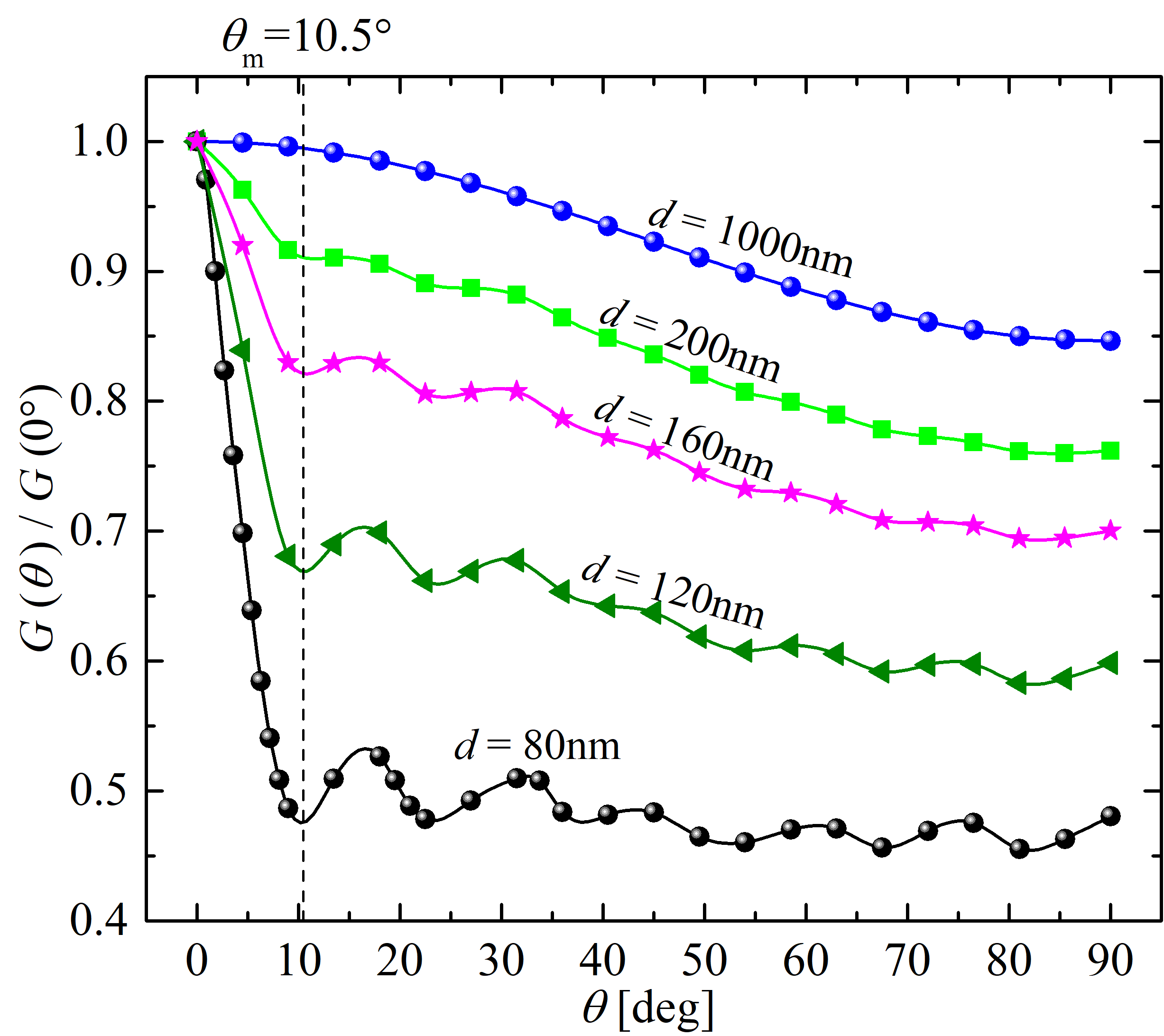}}
\caption{The ratio $G(\theta)/G(\theta=0^\circ)$ for circular gratings. Five different seprations $d$ are considered, $d=80$ nm, 120 nm, 160 nm, 200 nm and 1000 nm, respectively. $a=20$ nm, $T=300$~K, $p=60$ nm, $e=10a=200$ nm and $D=1.8~ \mu$m.}
\label{separation_dependence}
\end{figure}

We move here to the study of the separation $d$ dependence of the thermal conductance all along the twisting. The ratio of the thermal conductance $G(\theta)$ to the thermal conductance $G(\theta=0^\circ)$ is shown as a function of the twisting angle $\theta$ and for different values of $d$ in Fig.~\ref{separation_dependence}. Circular gratings with $D=1.8~ \mu$m are considered here.

The decreasing behaviour of  $G(\theta)/G(\theta=0^\circ)$ with the twisting angle $\theta$ is significantly dependent on the distance. For the smallest distance $d= 80$~nm, $G(\theta)/G(\theta=0^\circ)$ decreases from 1 to 0.47 and then oscillates in a pronounced way with increasing angles. As $d$ increases, the oscillations of $G(\theta)/G(\theta=0^\circ)$ with increasing the $\theta$ decreases gradually. Particularly, for a large $d$, $G(\theta)/G(0^\circ)$ decreases with $\theta$ monotonically and smoothly, which is similar to the monotonic and smooth decreasing dependence of near-field heat transfer coefficient $h/h_{\rm max} {\rm~(W/m^2)}$ with twisting angle observed for the 1D semi-infinite gratings \cite{Liu2014IJHMT,He2020IJHMT}. For a large $d$, $G(\theta)/G(\theta=0^\circ)$ varies in a much shorter range (e.g., for $d=200$~nm, it ranges from 1.0 to 0.8) as compared to the case with smaller separations (e.g., for $d=80$~nm , it ranges from 1.0 to 0.47). It's worthwhile to mention that the first-valley $\theta_{\rm m}$ is independent on the separation distance $d$ between the two gratings and is significantly dependent on the size of the gratings.

\begin{figure} [htbp]
\centerline {\includegraphics[width=0.48\textwidth]{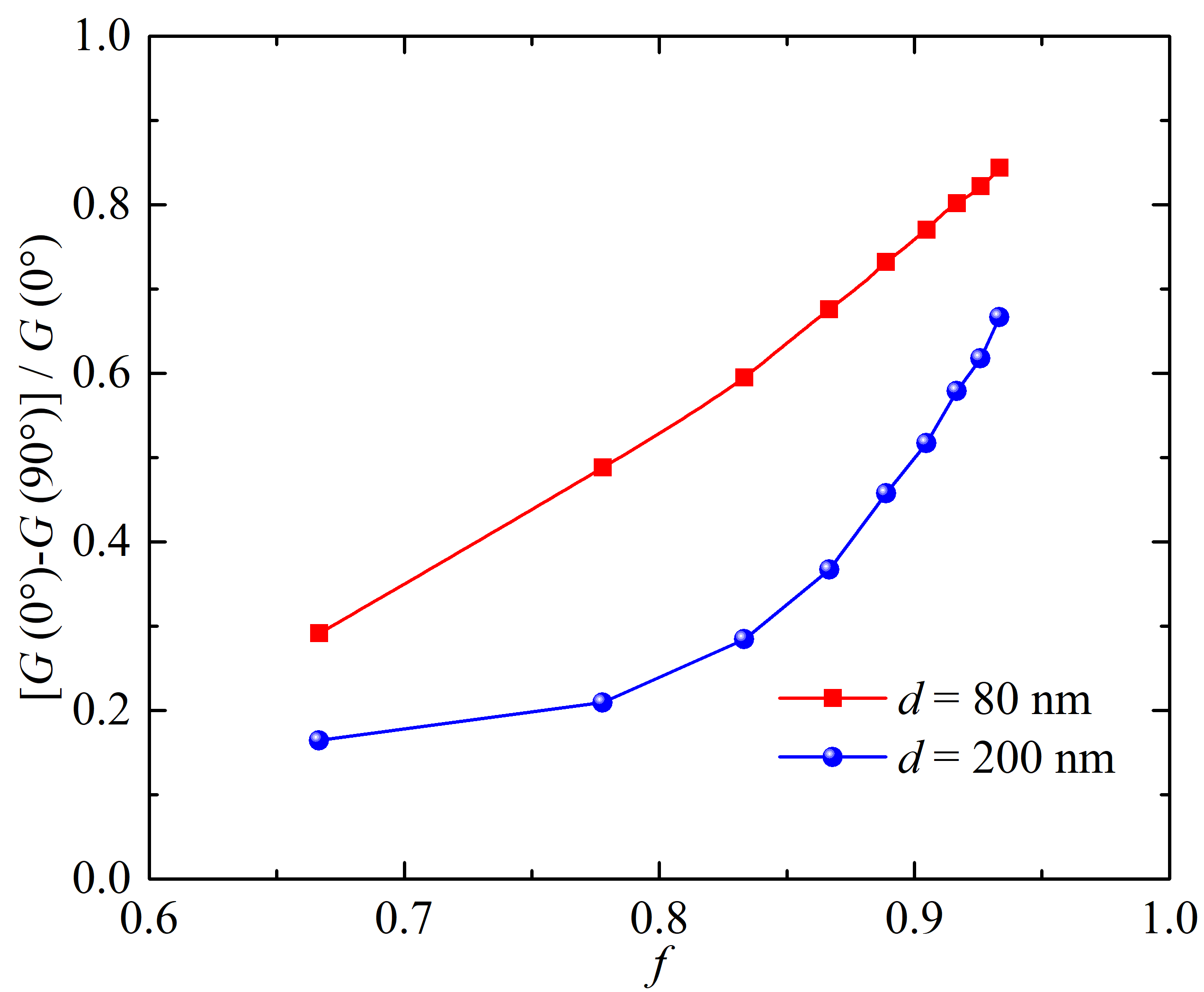}}
\caption{$[G(0^\circ)-G(90^\circ)]/G(0^\circ)$ as a function of the vacuum filling fraction $f$ at two different distances $d=80$ nm and $d=20$0 nm, and for circular gratings. $T=300$ K, $a=20$ nm, $p=60$ nm and $D=4~\mu$m.}
\label{modulation_rate}
\end{figure}

The vacuum filling fraction $f=(e-2a)/e$ is another parameter which can significantly influence the NFRHT. The modulation rate $[G(0^\circ)-G(90^\circ)]/G(0^\circ)$ is shown in Fig.~\ref{modulation_rate} as a function of $f$. $p=60$~nm. Two different distances $d=80$ nm and $d=20$0 nm are considered.

For the two different separations, the modulation rate $[G(0^\circ)-G(90^\circ)]/G(0^\circ)$ increases with the vacuum filling fraction $f$. The modulation rate for a small distance $d=80~$nm is always much larger than that of the case with a big distance $d=200~$nm because of a stronger coupling at small separations. The coupling between the two gratings with twisting angle is distance dependent.

In summary, the modulation of NFRHT between nanoparticle gratings by twisting is analysed beyond the EMA and infinite-size approximations, which is needed for realistic experimental analysis. We studied different finite-size grating shapes (square and circular) which present remarkably different features. When twisting one grating, characteristic oscillations of the thermal conductance $G(\theta)$ are observed (more pronounced for circular gratings), and fully due to finite-size and beyond-EMA effects. $G(\theta)$ converges to a given shape with increasing the size of gratings. In addition, the modulation rate $[G(0^\circ)-G(90^\circ)]/G(0^\circ)$ is significantly dependent on the vacuum filling fraction. This work may help for energy management at the nanoscale with relevant implications for MEMS and NEMS devices.

As for the infinite system, most recently, the Fourier modal method was raised for the description of nanoparticle lattices in the dipole approximation \cite{Fradkin2019,Fradkin2020}, which makes the direct investigation on NFRHT between infinite periodic nanoparticle lattices possible.

\begin{acknowledgments}
The support of this work by the National Natural Science Foundation of China (No. 51976045) is gratefully acknowledged. M.A. acknowledges support from the Institute Universitair\'e de France, Paris, France (UE). M.G.L. also thanks for the support from China Scholarship Council (No.201906120208).
\end{acknowledgments}

The data that support the findings of this study are available from the corresponding author upon reasonable request.


\providecommand{\noopsort}[1]{}\providecommand{\singleletter}[1]{#1}%

\end{document}